\documentclass[twoside,twocolumn,english,aps,showpacs,prl,superscriptaddress]{revtex4-1}

	\usepackage[usenames,dvipsnames]{xcolor}
	\usepackage{amsmath}
	\usepackage{amsfonts}
	\usepackage{amssymb}
	\usepackage[colorlinks=true,citecolor=blue,linkcolor=red]{hyperref}
	\usepackage{graphicx}
	\usepackage{bbold}					
	\usepackage[makeroom]{cancel}		
	\usepackage{multirow}				
	\usepackage[normalem]{ulem}        
	\usepackage{array}
	\usepackage{comment}
	\usepackage{makecell}
	\usepackage{pdfpages}
	\makeatletter
	\AtBeginDocument{\let\LS@rot\@undefined}
	\makeatother

\makeatletter

\makeatletter
\newcommand*{\rom}[1]{\expandafter\@slowromancap\romannumeral #1@}
\makeatother

\usepackage{times}{\Large }

\makeatother

\usepackage{babel}

\begin{document}

\preprint{APS/123-QED}

\title{Double Fu-teleportation and anomalous Coulomb blockade in a Majorana-hosted superconducting island}

\author{Yiru Hao}
\thanks{These authors contributed equally to this work.}
\affiliation{State Key Laboratory of Low Dimensional Quantum Physics, Department of Physics, Tsinghua University, Beijing, 100084, China}

\author{Gu Zhang}
\thanks{These authors contributed equally to this work.}
\affiliation{Beijing Academy of Quantum Information Sciences, Beijing 100193, China}

\author{Donghao Liu}
\affiliation{State Key Laboratory of Low Dimensional Quantum Physics, Department of Physics, Tsinghua University, Beijing, 100084, China}

\author{Dong E. Liu}
\email{Corresponding to: dongeliu@mail.tsinghua.edu.cn}
\affiliation{State Key Laboratory of Low Dimensional Quantum Physics, Department of Physics, Tsinghua University, Beijing, 100084, China}
\affiliation{Beijing Academy of Quantum Information Sciences, Beijing 100193, China}
\affiliation{Frontier Science Center for Quantum Information, Beijing 100184, China}

\date{\today}

\begin{abstract}
We study the temperature dependence of Coulomb Blockade peak conductance based on a Majorana-hosted superconducting island. In the low-temperature regime, we discover a coherent double Fu-teleportation (FT) process, where any independent tunneling process always involves two coherent FTs; and we also find an anomalous universal scaling behavior, which shows a crossover from a $[\max(T,eV)]^6$ to a $[\max(T,eV)]^3$ conductance behavior as increasing energy scale. In the high-temperature regime, using the familiar rate equation method, we find that the conductance is proportional to the reciprocal of the temperature and shows a non-monotonic temperature-dependence. Both the anomalous power-law behavior and non-monotonic temperature-dependence can be distinguished from the conductance peak in the traditional Coulomb block, and therefore, serve as a hallmark for the non-local transport in the topological superconducting island.
\end{abstract}

\maketitle


\textbf{\em Introduction.} The experimental search for Majorana zero modes (MZMs)~\cite{ReadGreen,Kitaev:2001} is a promising yet hotly debated topic in recent years. The standard tunneling spectroscopy detection has not yet reached the robust quantized zero bias value $2e^2/h$~\cite{sengupta2001midgap,Law09} in nanowire~\cite{Mourik2012,deng2012anomalous,das2012zero,finck2013anomalous,churchill2013superconductor,deng2016majorana,ZhangNC2017,chen2017experimental,suominen2017zero,nichele2017scaling,ZhangNN2018,vaitiekenas2018effective,de2018electric,bommer2019spin,grivnin2019concomitant,FrolovNP2021,FrolovScience2021,pan2020arXiv,zhang2021large,song2021large} or vortex Majorana platforms~\cite{DHong-2018-Sci,FDongLai-2018-PRX,KLingYuan-2019-NatPhys,Hanaguri-2019-NatM}. The dissipative tunneling scheme~\cite{liu2013filter,ChristianPRB20,liu2021universal,zhang2021suppressing} provides a tool to 
distinguish the local Majorana resonance from trivial signals using interaction-induced quantum phase transition/quantum criticality.
Instead of the single-terminal measurement where only local state is probed, the two-terminal measurement can capture the non-local feature of the topological island~\cite{Fu2010PRL}. It thus provides more confirmative and direct evidence of the presence of MZMs.
In addition, this non-local feature is directly associated with the topological protection of the potential Majorana-based quantum information processing. 
This fact has inspired multiple recent theoretical~\cite{Fu2010PRL,Heck2016PRB,Lutchyn2017PRL-Island,Chiu2017PRB,Pikulin2019PRL,LiuCX2019PRAP,LiuDH2020PRB,Lai2021PRB} proposals and experimental~\cite{Albrecht2016N,Farrell2018PRL,Shen2018NC,Vait2020Science,WhiticarNC2020-coherent-MajoranaIsland,LarsenPRL2020-Parity,Shen2021PRB,wang2021supercurrent,poschl2022nonlocal,sabonis2021comparing} efforts.




Confined quantum islands usually feel an electrostatic energy, and therefore, the electron transport shows Coulomb blockade (CB) signatures with conductance oscillations ~\cite{glazman2005review}. 
In the presence of superconductivity (SC), the signature of CB is modified.
When the order parameter is larger than the charging energy, the single electron (or $1e$) tunneling is suppressed and only the $2e$ cooper pair tunneling survives, leading to the oscillation with $2e$ periodicity~\cite{hekking1993CBinSC}.
This $2e$-feature maybe however not the case when facing a topological SC island.
Indeed, the non-local transport through a topological SC island~\cite{Fu2010PRL}, known as the Fu-teleportation (FT), has a $1e$ periodicity in CB.
Afterwards, a more careful theoretical analysis was carried out to obtain the CB signatures~\cite{Heck2016PRB}.
We summarize three major features of the FT~\cite{Fu2010PRL,Heck2016PRB}
: 1) for all different cases, the CB peak height increases while lowering the temperature;
2) CB oscillations with $1e$ and $2e$ period respectively accompany the tunneling of $1e$ quasiparticles and $2e$ Cooper pairs,
and 3) The CB peak shape of FT is the same as that of a resonant level model~\cite{Heck2016PRB} captured by Breit-Wigner formula~\cite{Stone1985PRL,Beenakker1991PRB}. Because of these coincidences with the standard CB features, it is not yet known whether or not the two-terminal CB island could provide a hallmark for verifying MZMs.

\begin{figure}[t]
\centering
  \includegraphics[scale=0.4]{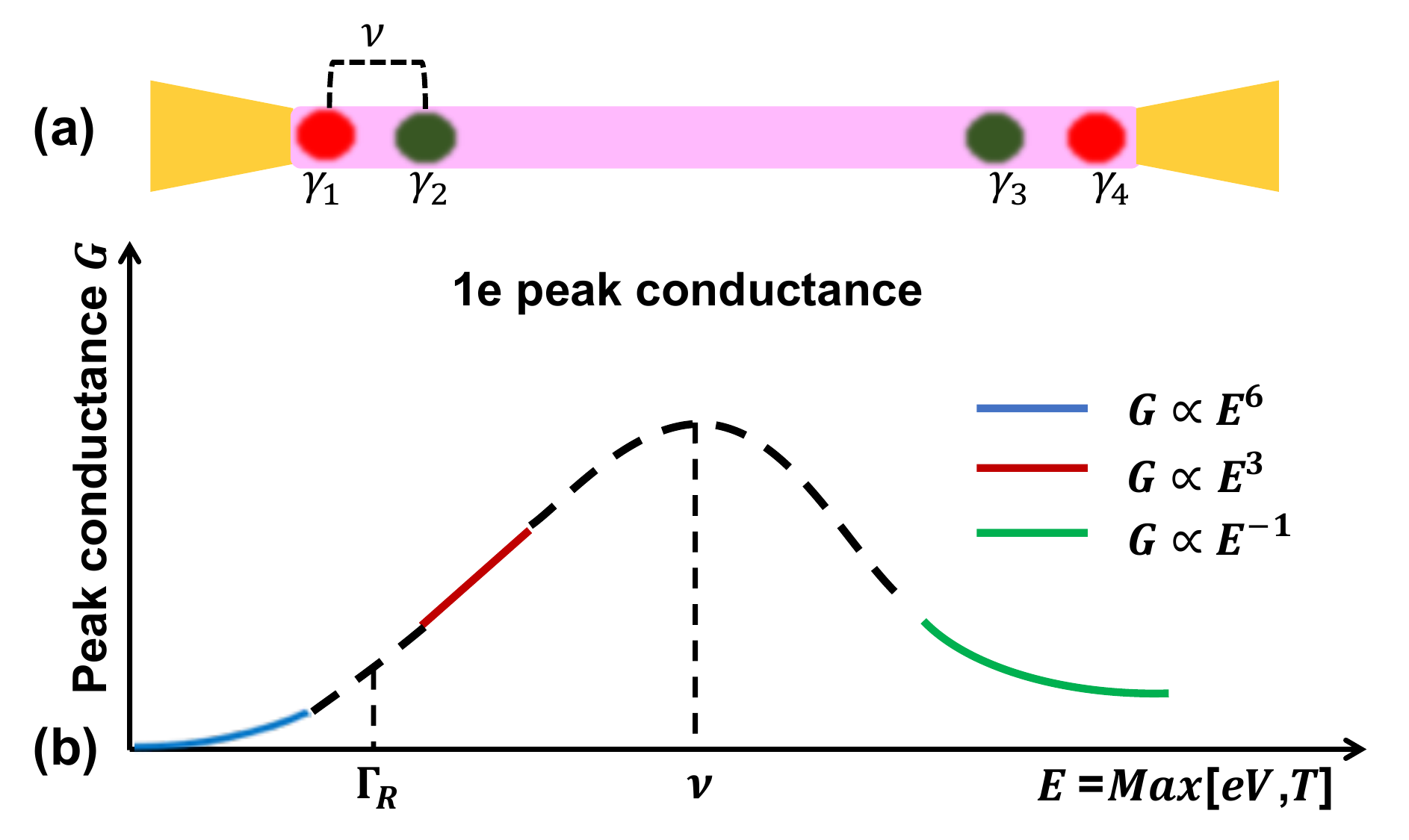}
  \vspace*{-0.3cm}
\caption{\label{fig:one} (a) Illustration of our system that consists of four MZMs on opposite sides of the nanowire. Two of them ($\gamma_1$ and $\gamma_2$) are coupled with the strength $\nu$, and the others ($\gamma_3$ and $\gamma_4$) are either decoupled MZMs or quasi-MZMs.
(b) The CB peak conductance through our Majorana-hosted island that is tuned to the half filling.
}
\vspace*{-0.3cm}
\end{figure}

\textbf{{\em Main results.}}
In this work, we study the two-terminal transport through a CB island that hosts a MZM and two coupled MZMs at opposite sides of the island [Fig.\ref{fig:one}(a)].
Based on our analysis, when $\nu$ is much larger than the the MZM-lead coupling $\Gamma_R$, such a Majorana-hosted SC island displays unique features. As the starter, the $1e$ conductance peak locations are independent of the value of $\nu$~\cite{supmat}. This is in stark contrast to the FT where the peak position is inter-MZM coupling dependent~\cite{Heck2016PRB}.
More interestingly, our system is expected to display a non-monotonic temperature dependence at the $1e$ CB conductance peak [Fig.\,\ref{fig:one}(b)]. In the lowest energy regime, we predict a coherent double FT with the peak conductance scaling $\sim[\max(T,eV)]^6$, where any tunneling event connecting two leads involves two coherent FT processes.
When energy increases (above the level broadening), the paired FTs lose coherence and the conductance crossover to the $\sim [\max(T,eV)]^3$ scaling.
Further increasing the energy, the $1e$ peak height reaches its maximum when the energy is around the inter-MZM coupling $\nu$.
Above this energy, the $1e$ peak height starts to decrease and approaches the standard FT results~\cite{Heck2016PRB} with the $\sim 1/T$ scaling.
Both the anomalous temperature dependence and the coherent double FT effect associated with the $1e$ CB peak can be used as hallmarks for Majorana-assisted non-local transport, as they are in sharp contrast to those of the normal CB systems.

\textbf{{\em Model and low-energy effective theory.}}
One possible realization of the proposed system is shown in Fig.~\ref{fig:one}(a), where a floated superconductor-proximitized nanowire (the pink line) weakly couples to one normal lead at each side. Under the protection of the smooth potential~\cite{Kells2012PRB}, two pairs of partially separated MZMs (or quasi-Majoranas) emerge at two ends of the nanowire in the (topologically) trivial regime~\cite{Kells2012PRB,Roy2013PRB,Cayao2015PRB,Pablo2016SR,Liu2017PRB,Setiawan2017PRB,Fernando2018PRB,reeg2018PRB,Moore2018PRBTwo,Moore2018PRBQuantized,Vuik2019SP,Awoga2019PRL,Cao2019PRL,pan2020PRR,pan2020PRB}.
In Fig.~\ref{fig:one}(a), we model our system with four quasi-Majoranas at each end as $\gamma_1$, $\gamma_2$, $\gamma_3$ and $\gamma_4$.
With these Majorana operators, we construct two independent auxiliary fermionic operators $d_1=(\gamma_1+i \gamma_4)/2$ and $d_2=(\gamma_2+i \gamma_3)/2$. We tune the left tunneling barrier into a steep shape to partially overlap $\gamma_1$ and $\gamma_2$ with the coupling strength $\nu$, and consider the regime that only $\gamma_1$ of the pair effectively coupled to the left lead and the $\gamma_2-$ lead coupling is exponentially suppressed. In addition, we keep the right barrier in a shallow shape to make sure the coupling between the other pair is negligible~\cite{Vuik2019SP}. We can also consider the setup with both a coupled MZM pair and a single MZM in a regular Majorana-hosted island.


For the proposed Majorana-hosted island system [shown in Fig.~\ref{fig:one}(a)], the total Hamiltonian can be written as
\begin{equation}\label{eq:one}
H=H_\text {lead}+U_c+H_\text{coupling}+H_\text{T},
\end{equation}
where $H_\text{lead}=\sum_{k, j=L,R} \epsilon_{j}(k) c_{j, k}^{\dagger} c_{j, k}$ describes two non-interacting leads. $U_c=E_c(N-n_g)^2$ is the electrostatic energy induced by the Coulomb interaction between electrons in the nanowire island. $E_c$ is the charging energy which is smaller than the proximity SC gap but larger than other relevant energy scales. $N$ represents the total number of electrons, and $n_g$ is tunable through a backgate voltage. $H_\text{ coupling }=i \nu \gamma_{1} \gamma_{2}$ is the coupling term between $\gamma_{1}$ and $\gamma_{2}$.
As $\gamma_1$ and $\gamma_2$ are both close to the left lead [Fig.~\ref{fig:one}(a)], their coupling $\nu$ does not change the conductance peak position (i.e., $n_g = 2n_0 + 1/2$, where $n_0$ indicates the number of hosted Cooper pairs).
This is in stark contrast to the $\nu$-dependent peak position of a Fu-teleportation, where the inter-MZM coupling is between two non-local MZMs through which the non-local transport is realized.
Neglecting the contribution of the quasi-particle states above the SC gap to the electric current at low energies, the tunneling Hamiltonian is
\begin{equation}\label{eq:two}
H_{T}\!=\!\lambda_{L}\sum_{k,L} c_{kL}^{\dagger}\gamma_1e^{-i\varphi/2}\!+\!\lambda_{R}\sum_{k,R} c_{kR}^{\dagger}\gamma_4e^{-i\varphi/2}\!+\!h.c.,
\hspace{-0.2em}
\end{equation}
where $\lambda_{L,R}$ denotes the respective tunnel matrix elements, and $e^{\pm i\varphi/2}$ raises/lowers $N$ by one charge unit~\cite{BeriPRL2013}.

Due to the Coulomb blockade, we can further map the model to its low-energy sector. With $n_g$ a half integer ($n_g=2n_0+1/2$), we only need to consider states in the Hilbert space $\{|00\rangle,|10\rangle,|11\rangle,|01\rangle\}$ spanned by basis vectors that dominate low-energy current tunneling, where $|i,j\rangle$ refers to the state with particle numbers $i$ and $j$ respectively for $d_1$ and $d_2$.
To further explore the relevance to Fu-teleportation~\cite{Fu2010PRL}, we define two impurity operators: one fermionic $f_{1} = |00\rangle\langle 10|-| 11\rangle\langle 01| = (d_1 - d_1^\dagger)\exp(-i\varphi/2)$ and one bosonic $f_{2} =|00\rangle\langle 11|-| 10\rangle\langle 01| = -d_1 d_2 - d_1^\dagger d_2$.
They are independent since $[ f_1, f_2 ] = \left\{ f_1, f_2 \right\} = 0$.
The bosonic operator $f_2$ is equivalent to a spin operator, via the mapping $f_2 = S_-$, $f_2^{\dagger}  = S_+$, and $S_z =  f_2^{\dagger} f_2 - 1/2$~\cite{supmat}. With analysis above, for the peak positions (i.e., half-filling $n_g = 1/2$), the effective Hamiltonian becomes
\begin{equation}\label{eq:six}
\begin{split}
H_\text{eff}=& H_\text{leads} - 2 \nu S_y -2 \lambda_L \sum_{k} c^{\dagger}_{kL} S_z f_1 \\
&+\lambda_R \sum_k c^{\dagger}_{kR}  f_1 + h.c.,
\end{split}
\end{equation}
where we have used the fact that $S_y =i (-S_+ + S_-)/2$. 

It is instructive to study the equilibrium conductance behavior in Eq.~\eqref{eq:six} at zero temperature.
The impurity Hamiltonian $- 2 \nu S_y$ has its ground state $|G \rangle=(-i, 1)^{T}$ which has a zero $S_z$ expectation $\langle G|S_z|G \rangle=0$. Consequently, the island tunneling to the left lead vanishes at zero-energies ($T = eV = 0$), leading to a zero conductance at the low-energy fixed point. This result can be understood that the influence of the coupling term is to form a localized Andreev bound state that prevents non-local tunneling completely at zero energies.

\textbf{\em Double Fu-teleportation at low-Temperature.}
Let us first analyze the fluctuations near the low-energy fixed point of the effective Hamiltonian Eq.~\eqref{eq:six} using the leading irrelevant operator.
Eq.~\eqref{eq:six} tells us that the tunneling at the left lead $\propto \lambda_L$ changes the impurity between the low-energy and the high-energy states. This is classically forbidden when $\nu \gg \max(T,eV),\Gamma_R$, as the energy of the high-energy state is unaffordable by either thermal ($\sim T$), quantum ($\sim\Gamma_{R} = \pi \rho |\lambda_{R}^2|$, where $\rho$ refers to the lead density of states) fluctuations, or the non-equilibrium driving ($\sim eV$).
Quantum mechanically, however, tunneling is possible via
high-order tunneling operators that transport particles through high-energy virtual states.
More specifically, when $f_1$ is occupied, we can construct a higher-order tunneling operator with three sub-operators: (i) $c^{\dagger}_{qL} S_z f_1$, (ii) $f_1^\dagger c_{kR}$ and (iii) $c^\dagger_{pL} f_1 S_z$. Each operator alone is forbidden at low energies due to the energy penalty.
However, if high-energy states occur virtually, these operators together combine into a higher-order operator $c^{\dagger}_{p L} S_z f_1 \cdot f_1^{\dagger} c_{k R} \cdot c^{\dagger}_{q L} S_z f_1$ (labeled as \textbf{\textit{process A}}) that bridges two energy-allowed real states.
To produce a persistent current, process A is followed by the operator $c_{kR} f_{1}^{\dagger}$ that returns the island to its initial state (labeled as \textbf{\textit{process B}}).
The successive occurrence of processes A and B leads to a persistent electron transport from the right to the left lead.
Noteworthily, one needs a careful treatment of the operator $\mathcal{O}_A$ of process A, since it involves two fermionic operators in the left leads. Indeed, after a careful Schiffer-Wolff transformation~\cite{hewson1997kondo,supmat}, the process A operator
\begin{equation}
    \mathcal{O}_A\! = \!\sum_{\epsilon_p>\epsilon_q,k} \! \frac{2 (\epsilon_{p} - \epsilon_{q})}{\nu^3} \lambda_L^2 \lambda_R c^{\dagger}_{p L} S_z f_1 \cdot f_1^{\dagger} c_{k R} \cdot c^{\dagger}_{q L} S_z f_1
\label{eq:oa}
\end{equation}
contains a momentum-dependent prefactor, and a conditional summation $\epsilon_p>\epsilon_q$.
This prefactor vanishes in zero-energy (i.e., zero-temperature and in-equilibrium) situations where $\epsilon_p = \epsilon_q = 0$. For finite-energy situations, $(\epsilon_p - \epsilon_q)^2 \sim [\max(T,eV)]^2$ after the summation over momenta.

In this low-energy situation, the effective transmission rate \cite{Ingold1992SUS} becomes $\tau_{\text{seq}} \equiv 1/\Gamma_\text{seq} = 1/\Gamma_{L}^\text{eff} + 1/\Gamma_R$.
where $\Gamma_{L}^\text{eff}=\pi \rho |\langle S_z \rangle^{2}||\lambda_{L}^2|$ refers to the effective level broadening from the higher-order operator.
Following a standard renormalization group (RG) analysis~\cite{AltlandSimonsBook}, $\Gamma_L^\text{eff}$ is RG irrelevant and becomes increasingly unimportant at low energies, in comparison to the level broadening $\Gamma_R$ of the RG-relevant right-lead coupling.
With low-enough energies, $\Gamma_R \gg \Gamma_{L}^\text{eff}$, and the sequential tunneling rate $\tau_\text{seq} \approx 1/\Gamma_{L}^\text{eff}$ is almost determined by the effective level broadening $\Gamma_{L}^\text{eff}$ of the higher-order operator $\mathcal{O}_A$.
With this knowledge in mind, we begin to analyze the system low-energy conductance features in two limiting cases.

\begin{figure}[t]
\centering
  \includegraphics[scale=0.5]{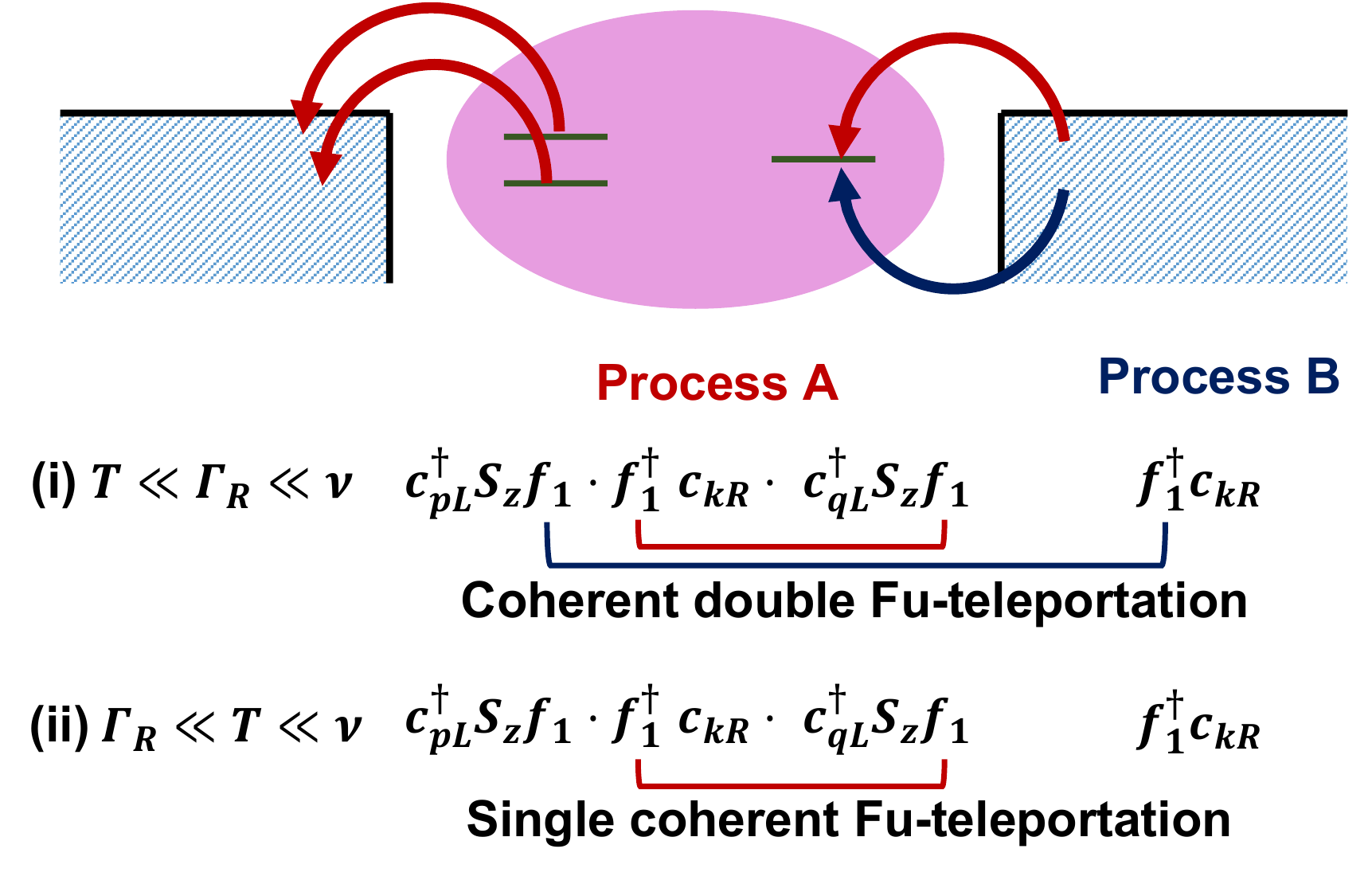}
\caption{\label{fig:two} The schematic diagrams of high-order coherent operators. Process A (red arrow) and B (blue arrow) are coherent in the extremely low-temperature regime $T\ll \Gamma_\text{seq} \ll \nu$, where they together construct the coherent double Fu-teleportation (FT). In the regime $\Gamma_\text{seq} \ll T \ll \nu$, these two processes decohere, and only single coherent FT exists.}
\end{figure}

In the extremely low energy regime $\max(T,eV) \ll \Gamma_R \ll \nu$, processes A and B are coherent, leading to a \textbf{\textit{coherent double Fu-teleportation}} as shown in Fig.~\ref{fig:two}(a).
Indeed, in this regime, an A or B process alone is forbidden as they relax the $f_1$-right lead hybridization, leading to an energy penalty $\Gamma_R$ that is unaffordable by the fluctuation $\max(T,eV) \ll \Gamma_R$.
Consequently, A and B processes always occur coherently.
This coherent double Fu-teleportation can be captured experimentally via the low-energy current measurement.
Indeed, the impurity operator $f_1$ becomes dynamical in this regime~\footnote{Similar treatment has been taken, e.g., for Kondo~\cite{Affleck1991} and two-impurity Kondo~\cite{SelaAffleckPRL09} systems.}, and $\mathcal{O}_A$ of Eq.~\eqref{eq:oa} now effectively consists of six non-interacting lead fermions.
Operator $\mathcal{O}_A$ then has the scaling dimension $\alpha= 6 \times 1/2 = 3$ at low energies, which equals six times that of one free fermionic operator (i.e., $ 1/2$~\cite{cftBook}).
This scaling dimension indicates the suppressed tunneling $\sim [\max(T,eV)]^{2(\alpha - 1)} = [\max(T,eV)]^4 $ at low energies~\cite{Nazarov1992charge,KaneFisher2}.
This fact, in combination with the extra power from the prefactor of $\mathcal{O}_A$ in Eq.~\eqref{eq:oa}, leads to the expected low-energy conductance $G \propto [\max(T, eV)]^6$, which is anomalous and highly distinguishable from conductance features through normal structures. This high power-law in energy is a strong signature of non-local coherent tunneling. Indeed, the energy-forbidding of local tunneling operators reveals the higher order non-local events that manifest the deep inner structures of the system.

This anomalous conductance feature accompanies the crossover to another feature for the regime with a slightly higher temperature $\Gamma_R \ll T \ll \nu$. 
In this regime, the lead-$f_1$ hybridization is relaxed (after which $f_1$ loses its dynamics), thus allowing the individual occurrences of A and B [Fig.~\ref{fig:two}(b)]. 
Now, the operator $\mathcal{O}_A$ has the scaling dimension $\alpha = 3/2$ (three times that of a free fermion $c_k$), indicating the low-energy power law $\sim \max(T,eV)$~\cite{KaneFisher1}. Once again, we combine this power law with that from the prefactor of $\mathcal{O}_A$, leading to the conductance feature $G \sim [\max(T,eV)]^3$ for low energies.

These two anomalous conductance power laws are among the central points of our work.
Briefly, we anticipate the crossover between these power laws in the low-energy regime $\max(T,eV) \ll \nu$: When $\max(T,eV) \gg \Gamma_R$, the conductance is determined by operator $\mathcal{O}_A$, with $G \sim [\max(T,eV)]^3$; When energy decreases, $\mathcal{O}_A$ is modified by the impurity-right lead coupling, and the related conductance feature crosses over to another power law $G \sim [\max(T,eV)]^6$ when finally $\max(T,V) \ll \Gamma_R$.
Both the anomalous power laws and the crossover over between them are highly exceptional, and thus capable in the experimental identification of the non-local teleportation.


To support our analysis, we calculate the low-bias conductance of our system at zero temperature using Green function technique~\cite{supmat}.
During our calculation, we treat the effective Hamiltonian exactly, except for $\mathcal{O}_A$. Indeed, as $\mathcal{O}_A$ is RG irrelevant, it is safe to treat $\mathcal{O}_A$ perturbatively to the leading order, where the current becomes
\begin{equation}
I = \frac{2e^2}{h}\int_{-\infty}^\infty dt e^{i e V t/\hbar} \Big\langle [ \mathcal{O}_A^\dagger (t), \mathcal{O}_A (0) ] \Big\rangle.
\label{eq:current_integral}
\end{equation}
In Eq.~\eqref{eq:current_integral} we have taken the trick (see, e.g., Refs.~\cite{KaneFisher1,ChamonFreedWenPRB96}) to deal with the bias as a time-dependent phase factor: by doing so, the correlation can be evaluated as if the system was in equilibrium.
\begin{figure}[t]
\centering
  \includegraphics[width=0.9 \columnwidth]{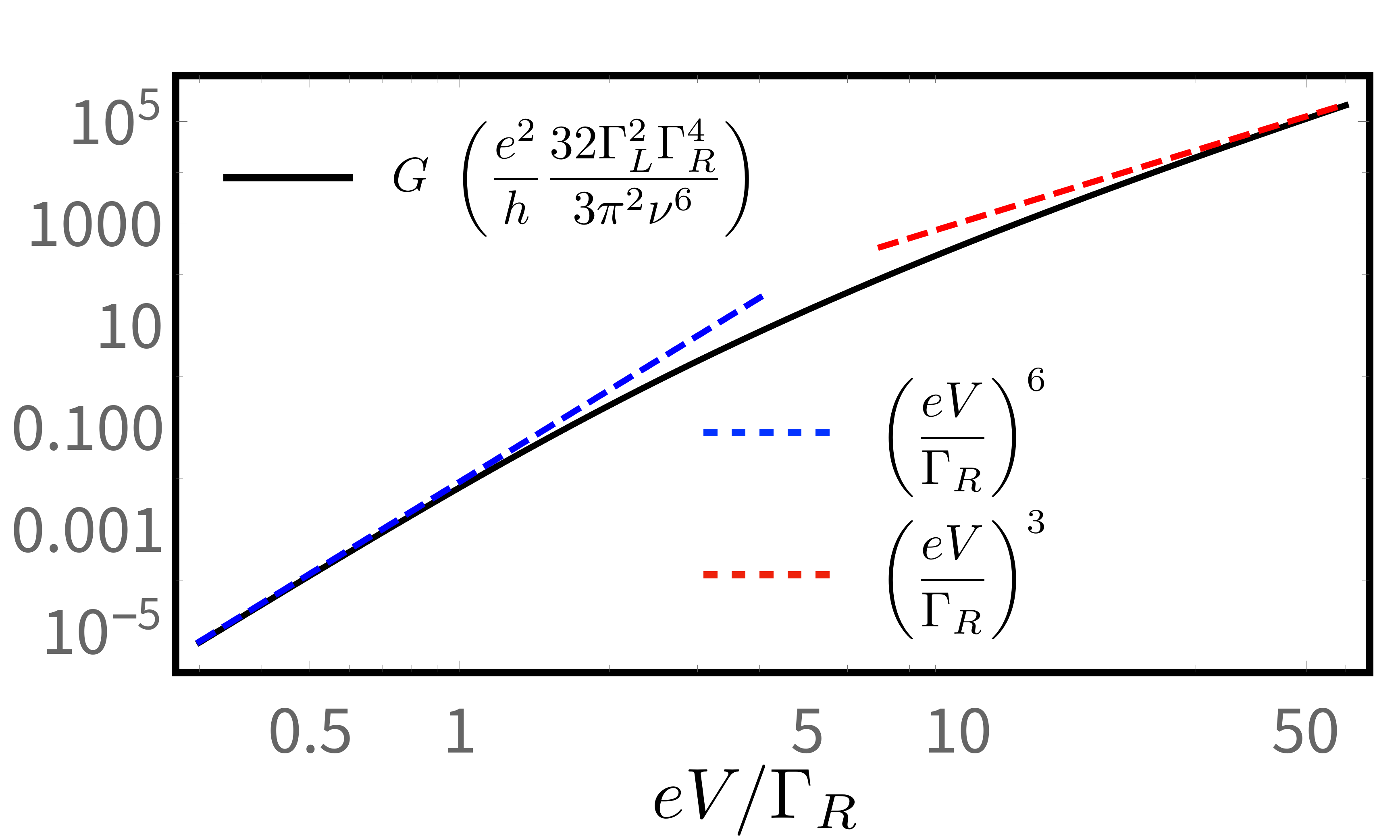}
  \vspace*{-0.3cm}
\caption{Conductance calculated with Eq.~\eqref{eq:current_integral} for our system. The blue and red dashed lines highlight power laws in different limits. The conductance $G\ll e^2/h$ is required in both limits.
}
\label{fig:exact_result}
\vspace*{-0.3cm}
\end{figure}
The current calculation is tedious but rather straightforward, with which we obtain the exact curve~\cite{supmat} shown in Fig.\,\ref{fig:exact_result}.
For two limiting cases, we can show that the conductance yields
\begin{align}
    &G \approx \frac{e^2}{h} \frac{4 \Gamma_L^2}{45\pi^2 \nu^6 \Gamma_R^2} (eV)^6 \propto (eV)^6, &\text{when $eV\ll \Gamma_R$,}\\
    &G \approx \frac{e^2}{h} \frac{16 \Gamma_L^2 \Gamma_R}{3\pi \nu^6} (eV)^3 \propto (eV)^3, &\text{when $eV \gg \Gamma_R$.}
\end{align}
These low-bias conductance power laws, valid in the regime $eV, \Gamma_R \ll \nu$, perfectly agree with our RG analysis above.

\textbf{{\em Single-electron tunneling in the high temperature regime.}}
In the high-temperature regime $\nu \ll T \ll E_{c}$, thermal fluctuation allows transport processes (e.g., $c^{\dagger}_{p L} S_z f_1$) that are otherwise forbidden in low-energies regimes.
It is then legitimate to evaluate the conductance
via the master equation formalism~\cite{Chiu2017PRB,Lai2021PRB,Heck2016PRB,glazman2005review}.
Of our case, the superconducting island contains four eigenstates, $|o_{1,2}\rangle=\left(\pm i|10\rangle+|01\rangle\right)/2$ and $|e_{1,2}\rangle=\left(\pm i|00\rangle+|11\rangle \right)/2$,
where $e$ and $o$ respectively label impurity states with even and odd parities. The occupation probability of each state follow the rate equations
\begin{equation}\label{eq:B1}
\begin{split}
\dot{P}_{\alpha}& =-\sum_{\beta} \Gamma_{\alpha \rightarrow \beta} P_{\alpha}+\sum_{\beta} \Gamma_{\beta \rightarrow \alpha} P_{\beta}, \\
\dot{P}_{\beta}&=-\sum_{\alpha} \Gamma_{\beta \rightarrow \alpha} P_{\beta}+\sum_{\alpha} \Gamma_{\alpha \rightarrow \beta} P_{\alpha}, \\
\end{split}
\end{equation}
where $P_{\alpha}$ and $P_{\beta}$ are the occupation probability of even $\alpha=|e_1\rangle,|e_2\rangle$ and odd $\beta=|o_1\rangle,|o_2\rangle$ parity states, respectively, and $\Gamma_{i \rightarrow f}=\Gamma_{i \rightarrow f}^{L}+\Gamma_{i \rightarrow f}^{R}=\sum_{j}\Gamma_{i \rightarrow f}^{j}$ represents the transition probability from state $|i\rangle$ to $|f\rangle$.
They can be evaluated from the Fermi golden rule
\begin{equation}\label{eq:B2}
\begin{aligned}
\Gamma_{\alpha \rightarrow \beta}^{j}=& \frac{2 \Gamma_j}{\hbar} \sum_{p} \delta\left(E_{\alpha}-E_{\beta}+\xi_{p}\right) f\left(\xi_{p}-\mu_{j}\right), \\
\Gamma_{\beta \rightarrow \alpha}^{j}=& \frac{2 \Gamma_j}{\hbar}  \sum_{p} \delta\left(E_{\beta}-E_{\alpha}-\xi_{p}\right)\left[1-f\left(\xi_{p}-\mu_{j}\right)\right],
\end{aligned}
\end{equation}
where chemical potentials $\mu_{L}=eV,\mu_{R}=0$, and $f(\epsilon)$ is the fermionic distribution. $E_{\beta}-E_{\alpha}$ is the energy difference between the odd $\beta$ and even $\alpha$ parity states, and $\xi_{p}$ is the electron energy in the leads. 

One can solve Eq.~\eqref{eq:B1} with the normalization requirement $\sum_{\alpha}P_{\alpha}+\sum_{\beta}P_{\beta}=1$.
With them, the current can be evaluated via
$I=e \sum_{\alpha,\beta} P_{\alpha} \Gamma_{\alpha \rightarrow \beta}^{L} -e \sum_{\alpha,\beta} P_{\beta} \Gamma_{\beta \rightarrow \alpha}^{L}$. At zero bias, the tunneling conductance becomes~\cite{supmat}
\begin{equation}\label{eq:B5}
G=\frac{e^{2}}{2T\hbar}\frac{\Gamma_{L} \Gamma_{R}}{\Gamma_{L}+\Gamma_{R}}\mathrm{\textrm{sech}}\left(\frac{\nu}{T}\right)^{2}\mathrm{\textrm{sech}}\left[\frac{E_{c}(1-2\delta_{g})}{2T}\right]^{2}.
\end{equation}
In agreement with our previous analysis, the conductance arrives at its peak value at half-filling $\delta_g = n_g - 2n_0 = 1/2$, independent of the inter-MZM coupling $\nu$.
As another feature, the peak conductance follows $\sim 1/T$ in the high-temperature $\nu \ll T \ll E_{c}$ limit, where the factor $\text{sech}(\nu/T)$ approximately equals one.
In the above calculation, the equilibration is reached from the self-consistent treatment of only the lead-island couplings. However, if the thermal effects of the island is mainly from the external environment, the island will first reach the thermal equilibrium. We call this situation "dirty" transport, and the conductance formula becomes slightly different~\cite{supmat}.

Combining the analysis in the low-energy regimes ($\max(T,eV) \ll \nu$) and the rate-equation calculations in the high energy regime ($\nu \ll \max(T,eV) \ll E_c$), we obtain the $1e$ conductance-peak features over the main energy regimes, as shown in Fig.~\ref{fig:one}(b).
Here the energy that induces the largest conductance is expected to be around $\max(T,eV)\sim \nu$, as given by the rate equation result Eq.~\eqref{eq:B5}.
Indeed, the semi-classical rate-equation is legitimate near this regime, where charge transport mainly relies on uncorrelated sequential tunnelings.
In the low-energy limit, conductance predicted by Eq.~\eqref{eq:B5} decays exponentially, instead of the polynomial feature predicted for coherent tunneling operators.
In this limit, one needs to go beyond the semi-classical picture, as coherent tunneling has become dominant.



\textbf{{\em Discussion.}}
 We mostly focus on the $1e$ CB conductance peak, i.e. $\delta_g=1/2$, of our Majorana-hosted SC island. We discover a novel double Fu-teleportation and anomalous Coulomb blockade, which manifest the deep inner structures of the system and could serve as a hallmark for the non-local transport in Majorana-hosted SC island. We emphasize that the analysis above is valid if $\nu \gg \Gamma_R$: otherwise the transport mimics that of a normal Fu-teleportation. In this sense, a crossover between the normal and anomalous conductance features is anticipated via the tuning of $\nu$ or $\Gamma_R$. For instance, if $\max(T,eV) \ll \Gamma_R \ll \nu$ initially, we anticipate to experimentally observe the crossover from the high-order power law feature $G \sim[\max(T,eV)]^6$ to a constant conductance via increasing the value of $\Gamma_R$.
We also emphasize that to observe these anomalous power laws and the crossover between them, the background zero-energy conductance $\sim \Gamma_{L,R}/E_c$ or $\sim \Gamma_{L,R}/\Delta_\text{sc}$ must be small, where $\Delta_\text{sc}$ refers to the superconducting gap.

When we tune the voltage to a different location $\delta_g=1$, electron states $N$ and $N+2$ are degenerate and form the $2e$ CB conductance peak~\cite{hekking1993CBinSC,Heck2016PRB}. We notice that the $2e$ peak height keeps almost constant in the relevant regime of this paper (i.e. $T\ll\Delta_{sc},E_C$). In addition, the $2e$ peak height is also very small compared to the $1e$ peak: for example, the ratio of the maximum of the 1e peak to the 2e peak is $\Delta_{sc}/(g T)$~\cite{Heck2016PRB} for the standard Fu-teleportation limit $\Delta_{sc}\gg T\gg \nu$, where $g\ll 1$ is the dimensionless tunneling conductance.

\begin{acknowledgments} 
\textbf{{\em Acknowledgments.}} Authors thank Zhan Cao and Xin Liu for helpful discussions. The work is supported by Natural Science Foundation of China (Grants No.~11974198) and Tsinghua University Initiative Scientific Research Program. 
\end{acknowledgments}

\bibliographystyle{apsrev4-1} 
\bibliography{DissipativeABS,refer,CB}

\onecolumngrid
\newpage


\section*{Supplementary material}

# Supplementary meterial for "Double Fu-teleportation and anomalous Coulomb blockade in a Majorana-hosted superconducting island"

In this supplementary information, we will provide details concerning: (I) The effective low-energy Hamiltonian of the impurity and the conductance peak position; (II) The effective low-energy tunneling Hamiltonian, (III) Detailed derivation of the current following Eq. (5) of the main text, (IV) High temperature regime: detailed derivation of Eq.(10) of the main text, and (V) High temperature regime for the case the thermal effect of the SC island is mainly from the external environment.

## I. EFFECTIVE LOW-ENERGY IMPURITY HAMILTONIAN AND CONDUCTANCE PEAK POSITION

In this section, we discuss the effective low-energy impurity Hamiltonian after neglecting the lead-impurity tunnelings $\lambda_L$ and $\lambda_R$. Of this situation, the Hamiltonian of the island becomes

$$U_c = E_c(N - n_g)^2 + \nu\gamma_1\gamma_2, \tag{S1}$$

where $n_g$ indicates the energetically most-preferred occupation number in the dot, and $\gamma_1$, $\gamma_2$ refer to two quasi-MZMs (or two coupled regular MZMs) next to the left lead. For a standard Fu-teleportation, the coupling between two non-local MZMs modifies the peak position (represented by the value of $n_g$) where two impurity states become energy-degenerate. Indeed, in Fu-teleportation, energies of these two impurity states are both $\nu$ and $n_g$-dependent. By contrast, in our system the peak position is instead $\nu$-independent.

To see this is indeed the case, we consider a small detuning $\delta n$ from the half-filling (i.e., $n_g = 2n_0 + 1/2 + \delta n$), with which the impurity Hamiltonian can be presented in the matrix form

$$H_{\text{impurity}} = \begin{pmatrix} E_c\delta n & 0 & 0 & -i\nu \\ 0 & -E_c\delta n & i\nu & 0 \\ 0 & -i\nu & -E_c\delta n & 0 \\ i\nu & 0 & 0 & E_c\delta n \end{pmatrix}, \tag{S2}$$

where matrix indices respectively represents impurity states $\{|00\rangle, |10\rangle, |01\rangle, |11\rangle\}$. After the exact diagonalization of $H_{\text{impurity}}$, we figure out its four eigenstates with their corresponding energies

$$\begin{aligned}
|\psi_1\rangle = |o_1\rangle &= \frac{1}{\sqrt{2}}\left(i|10\rangle + |01\rangle\right), & \epsilon_{o1} &= \nu - E_c\delta n, \\
|\psi_2\rangle = |e_2\rangle &= \frac{1}{\sqrt{2}}\left(-i|00\rangle + |11\rangle\right), & \epsilon_{e2} &= \nu + E_c\delta n, \\
|\phi_1\rangle = |o_2\rangle &= \frac{1}{\sqrt{2}}\left(-i|10\rangle + |01\rangle\right), & \epsilon_{o2} &= -\nu - E_c\delta n, \\
|\phi_2\rangle = |e_1\rangle &= \frac{1}{\sqrt{2}}\left(i|00\rangle + |11\rangle\right), & \epsilon_{e1} &= -\nu + E_c\delta n,
\end{aligned} \tag{S3}$$

where $e$ and $o$ respectively label impurity states with even and odd parities. Assuming that $\nu > 0$, states $|\phi_1\rangle$ and $|\phi_2\rangle$ respectively have lower energies in comparison to that of $|\psi_1\rangle$ and $|\psi_2\rangle$. Clearly, when $\nu \neq 0$, degenerate states are possible only when $\delta n = 0$, i.e., when $n_g = 2n_0 + 1/2$ is indeed tuned to the half-filling. This criteria is the same of degeneracy for both low-energy and high-energy regimes discussed in the main text. Since degeneracy provides extra options in the tunneling of particles, we anticipate the occurrence of the 1e tunneling peak of our model at half filling, independent of the inter-MZM coupling constant $\nu$. Briefly, this irrelevance of $\nu$ to the peak position grounds in the fact that $\gamma_1$ and $\gamma_2$ of our model are local in space. By contrast, two coupling MZMs of a common Fu-teleportation are responsible for the non-local transport.

## II. EFFECTIVE LOW-ENERGY TUNNELING HAMILTONIAN

After obtaining eigenstates and their corresponding energies of the island, we discuss the influence of tunneling amplitudes. Here we assume that $\delta n = 0$, and focus only on the tunneling at the peak position. We check the effect of lead-island tunneling

operators on the island states, leading to

$$\begin{aligned}
\gamma_1 e^{-i\varphi/2}|\psi_2\rangle = |\phi_1\rangle, \quad & \gamma_1 e^{-i\varphi/2}|\phi_2\rangle = |\psi_1\rangle, \\
\gamma_4 e^{-i\varphi/2}|\phi_1\rangle = |\phi_2\rangle, \quad & \gamma_4 e^{-i\varphi/2}|\psi_1\rangle = |\psi_2\rangle.
\end{aligned} \tag{S4}$$

Eq. (S4) shows us the effect of $\gamma_1 e^{-i\varphi/2}$ and $\gamma_4 e^{-i\varphi/2}$: $\gamma_4 e^{-i\varphi/2}$ changes the parity state while conserving the system energy; while $\gamma_1 e^{-i\varphi/2}$ changes the impurity state between the low-energy states ($|\phi_1\rangle$ and $|\phi_2\rangle$) and the high-energy ones ($|\psi_1\rangle$ and $|\psi_2\rangle$).

Eq. (S4) provides an alternative way in the understanding of the effective spin operators in Eq. (3) of the main text: here the spin operator $S_z$ models the change of energy (i.e., different $S_y$ eigenvalues) induced by the impurity-left lead coupling. Following Eq. (S4), we understand the zero-conductance at zero energies as the lack of energy to visit the high-energy states ($|\psi_1\rangle$ and $|\psi_2\rangle$). Indeed, the operator $c_{kL}^\dagger \gamma_1 e^{-i\varphi/2}$ is forbidden at zero energies. In contrast, $c_{kR}^\dagger \gamma_4 e^{-i\varphi/2}$ is allowed, as it does not change the system energy. However, it alone can not produce a persistent current. Instead, a persistent current requires the higher order tunneling operator introduced in the main text that connects the island and the left lead. At zero energies, these higher-order processes have zero amplitudes following Eq. (4) of the main text.

As another key feature, in Eq. (3) of the main text we model the energy change induced by the impurity MZM $\gamma_1$ via the definition of the impurity operator $f_2 = |00\rangle\langle 11| - |10\rangle\langle 01| = -d_1 d_2 - d_1^\dagger d_2$. This operator is mappable to spin operators, as it satisfies the commutators: (i) $f_2^2 = (f_2^\dagger)^2 = 0 = S_-^2 = S_+^2$; (ii) $[f_2^\dagger f_2 - 1/2, f_2] = -f_2$ and $[f_2^\dagger f_2 - 1/2, f_2^\dagger] = f_2^\dagger$. These commutation relations perfectly agree with $[S_z, S_+] = S_+$ and $[S_z, S_-] = -S_-$ of spin operators. After substituting the impurity operator $f_2$ by the corresponding spin operators, we have arrived at Eq. (3) of the main text.

### III. CALCULATION OF THE CURRENT

In this section, we provide details on the derivation of the zero-temperature current under an applied bias. As the starter, in the main text we have shown that the bare tunneling at the left side, i.e., $-2\lambda_L \sum_k c_{kL}^\dagger S_z f_1$ is energetically forbidden, as it connects two states with different energies. However, it can be used to construct tunnelings through virtual states. Indeed, with the Schrieffer-Wolff transformation [S1], one can construct the higher-order tunneling Hamiltonian

$$\begin{aligned}
& \mathcal{O}_A \sum_{p,q,k} \lambda_L c_{pL}^\dagger S_z f_1 \frac{1}{\nu - \epsilon_p} \lambda_R f_1^\dagger c_{kR} \frac{1}{\nu + \epsilon_q} \lambda_L c_{qL}^\dagger S_z f_1 \\
& \approx \sum_{\epsilon_p, \epsilon_q, k} \frac{2(\epsilon_p - \epsilon_q)}{\nu^3} \lambda_L^2 \lambda_R c_{pL}^\dagger c_{qL}^\dagger c_{kR} f_1,
\end{aligned} \tag{S5}$$

where in the second line we have expand to the leading order of lead-state energies $\epsilon_p$ and $\epsilon_q$. For non-equilibrium or finite-temperature situations, $\epsilon_p - \epsilon_q$ respectively has the order of bias and temperature. It thus adds to the energy power-laws from RG analysis.

We visit the current through the superconducting island via the Greens function technique. In this work, since the only non-quadratic $\mathcal{O}_A$ is RG-irrelevant operator, we treat it perturbatively, while solving the rest of the Hamiltonian exactly. By doing so, the current operator becomes

$$\begin{aligned}
\hat{I} &= -\partial_t [f_1^\dagger f_1 + \sum_k c_{kR}^\dagger c_{kR}] = -[H_{\rm T}, f_1^\dagger f_1 + \sum_{k'} c_{k'R}^\dagger c_{k'R}] \\
&= 2i \sum_{p>q,k} (t_{p,q} c_{pL}^\dagger c_{qL}^\dagger c_{kR} f_1 - h.c.) \equiv 2i(L - L^\dagger),
\end{aligned} \tag{S6}$$

where the minus sign is added to define the $R \to L$ current as positive, $t_{p,q} = 2(\epsilon_p - \epsilon_q)\lambda_L^2 \lambda_R/\nu^3$, and $L = \sum_{p>q} t_{p,q} c_{pL}^\dagger c_{qL}^\dagger c_{kR} f_1$.

We calculate the current at zero temperature $T$, under a bias $V$ that is applied to the right lead. As a famous trick (see e.g., Refs. [S2, S3]), one can deal with this bias with the transformation $c_{kR}^\dagger \to c_{kR}^\dagger \exp(ieVt)$, after which the calculation can be delta with as if the system was in equilibrium.

With this trick, to the leading order of $t_L$, current can be calculated as

$$I = 2\int_{-\infty}^\infty dt\, e^{ieVt} \langle [L^\dagger(t), L(0)] \rangle = 2\int_{-\infty}^\infty dt\, e^{iVt} \left\{ \langle L^\dagger(t) L(0) \rangle + \langle L(0) L^\dagger(t) \rangle \right\}. \tag{S7}$$

The first part of this calculation equals

$$\int_{-\infty}^{\infty} dt e^{ieVt} \sum_{p>q,k} \sum_{p'>q',k'} t_{p,q} t_{p',q'} \langle c_{qL}(t) c_{pL}(t) c_{p'L}^\dagger c_{q'L}^\dagger \rangle \langle f_1^\dagger(t) c_{kR}^\dagger(t) c_{k'R}(0) f_1(0) \rangle$$

$$= \sum_{p>q} \sum_{k,k'} \int_{-\infty}^{\infty} dt e^{i(eV-\epsilon_p-\epsilon_q)t} t_{p,q} t_{p',q'} [1-n_F(\epsilon_p)][1-n_F(\epsilon_q)]$$

$$\times \left[ G^<(f_1^\dagger, c_{k'R}, -t) G^<(c_{kR}^\dagger, f_1, -t) - G^<(f_1^\dagger, f_1, -t) G^<(c_{kR}^\dagger, c_{k'R}, -t) \right]$$

$$= \sum_{p>q} \sum_{k,k'} \int_{-\infty}^{\infty} d\omega' \frac{4\lambda_L^4 \lambda_R^2}{\nu^6} (\epsilon_p - \epsilon_q)^2 [1-n_F(\epsilon_p)][1-n_F(\epsilon_q)]$$

$$\times \left[ \tilde{G}^<(f_1^\dagger, c_{k'R}, \omega') \tilde{G}^<(c_{kR}^\dagger, f_1, -eV+\epsilon_p+\epsilon_q-\omega') - \tilde{G}^<(f_1^\dagger, f_1, \omega') \tilde{G}^<(c_{kR}^\dagger, c_{k'R}, -eV+\epsilon_p+\epsilon_q-\omega') \right],$$

(S8)

where $n_F$ refers to the fermi distribution function. At zero temperature $T=0$, $n_F(\epsilon) = \Theta(-\epsilon)$ equals the step function.

Lesser Greens functions of Eq. (S8) can be obtained via standard method [S4], with the result

$$\sum_k \tilde{G}^<(f_1^\dagger, c_{kR}, \omega) = \sum_k \tilde{G}^<(c_{kR}^\dagger, f_1, \omega) = 2\pi i \rho \frac{\lambda_R \omega}{\omega^2 + \Gamma_R^2} n_F(\omega)$$

$$\tilde{G}^<(f_1^\dagger, f_1, \omega) = \frac{2i\Gamma_R}{\omega^2 + \Gamma_R^2} n_F(\omega), \quad \sum_{k,k'} \tilde{G}^<(c_{kR}^\dagger, c_{k'R}, \omega) = 2\pi i \rho \frac{\omega^2}{\omega^2 + \Gamma_R^2} n_F(\omega).$$

(S9)

Notice that the lead lesser Greens function contains an extra power of energy $\sim \omega^2$. This factor reflects the hybridization of the impurity $f_1$ by the right lead when $\Gamma_R \gg \omega \sim eV$.

With Eq. (S9), the target integral Eq. (S8) becomes

$$\sum_{p>q} \frac{16\lambda_L^4 \Gamma_R^2}{\nu^6} (\epsilon_p - \epsilon_q)^2 \left[ \frac{\arctan(\delta\omega)}{\Gamma_R} - \frac{1}{\delta\omega} \ln\left( \frac{\delta\omega^2 + \Gamma_R^2}{\Gamma_R^2} \right) \right] \mathcal{S}_{p,q}.$$

(S10)

where $\delta\omega = eV - \epsilon_p - \epsilon_q$. Here the summation over $p, q$ is taken in the area

$$\mathcal{S}_{p,q} = \{ p, q \mid \epsilon_p > 0, \epsilon_q > 0, \epsilon_p + \epsilon_q < eV \text{ and } \epsilon_p > \epsilon_q \},$$

(S11)

which is a triangle in the $(\epsilon_p, \epsilon_q)$ space. The full expression of the conductance becomes

$$G = \frac{e^2}{h} \frac{32 \Gamma_L^2 \Gamma_R^2}{3\pi^2 \nu^6} \left\{ -8(eV)^2 + eV\Gamma_R \left[ 9 + \frac{(eV)^2}{\Gamma_R^2} \right] \arctan\left( \frac{eV}{\Gamma} \right) + \right.$$

$$\left. + \frac{(eV)^2}{\Gamma^2} \left[ -1 + 3\left( \frac{eV}{\Gamma} \right)^2 \right] \ln\left[ 1 + \left( \frac{eV}{\Gamma} \right)^2 \right] + \frac{3}{2}(eV)^2 \text{Li}_2\left[ -\left( \frac{eV}{\Gamma_R} \right)^2 \right] \right\},$$

(S12)

with which we plot Fig. 3 of the main text. Here $\text{Li}_n$ refers to the polylogarithm function. In two limiting cases, the conductance approximately becomes

$$G \approx \frac{e^2}{h} \frac{4\Gamma_L^2}{45\pi^2 \nu^6 \Gamma_R^2} (eV)^6, \quad \text{when } eV \ll \Gamma_R,$$

$$G \approx \frac{e^2}{h} \frac{16\Gamma_L^2 \Gamma_R}{3\pi \nu^6} (eV)^3, \quad \text{when } eV \gg \Gamma_R.$$

(S13)

## IV. 1E CONDUCTANCE OF THE CLEAN CASE IN $\nu \ll T \ll E_c$

The occupation probability of electrons at different energy levels can be described by rate equations

$$\dot{P}_\alpha = -\sum_\beta \Gamma_{\alpha \to \beta} P_\alpha + \sum_\beta \Gamma_{\beta \to \alpha} P_\beta,$$

$$\dot{P}_\beta = -\sum_\alpha \Gamma_{\beta \to \alpha} P_\beta + \sum_\alpha \Gamma_{\alpha \to \beta} P_\alpha,$$

(S14)

where $\dot{P}_\alpha$ and $\dot{P}_\beta$ are the occupation probability of even $\alpha = |e_1\rangle, |e_2\rangle$ and odd $\beta = |o_1\rangle, |o_2\rangle$ parity state, respectively, and $\Gamma_{i\to f} = \Gamma^L_{i\to f} + \Gamma^R_{i\to f} = \sum_j \Gamma^j_{i\to f}$ represents the transition probability between different impurity states $|i\rangle$ and $|f\rangle$. Assuming that the occupation probability of the single particle state in the lead follows the Fermi-Dirac distribution $f(\omega) = 1/(1+e^{\omega/kT})$, the following expressions of $\Gamma^j_{i\to f}$ are obtained from the Fermi golden rule:

$$\Gamma^j_{\alpha\to\beta} = \frac{2\Gamma_j}{\hbar}\int d\xi_p\, \delta\left(E_\alpha - E_\beta + \xi_p\right) f\left(\xi_p - \mu_j\right),$$
$$\Gamma^j_{\beta\to\alpha} = \frac{2\Gamma_j}{\hbar}\int d\xi_p\, \delta\left(E_\beta - E_\alpha - \xi_p\right) \left[1 - f\left(\xi_p - \mu_j\right)\right], \tag{S15}$$

where $\mu_L = eV, \mu_R = 0, \Gamma_j = \pi\rho|\lambda_j^2|$ describes transition amplitude, $\rho$ is the density of states, $E_\beta - E_\alpha$ is the energy difference between odd $\beta$ and even $\alpha$ parity states, and $\xi_p$ is the electron energy in the leads.

In the steady state, rate equations become $\dot{P}_\alpha = 0, \dot{P}_\beta = 0$. Combining the normalization conditions $\sum_\alpha P_\alpha + \sum_\beta P_\beta = 1$, we can work out the occupation probabilities of the four impurity states $P_{|e_1\rangle}$, $P_{|e_2\rangle}$, $P_{|o_1\rangle}$ and $P_{|o_1\rangle}$. Then the steady current can be easily calculated, the expression of the current is

$$I = e\sum_{\alpha,\beta} P_\alpha \Gamma^L_{\alpha\to\beta} - e\sum_{\alpha,\beta} P_\beta \Gamma^L_{\beta\to\alpha} \tag{S16}$$

From Eq. (S4), only two specific transitions $|e_1\rangle \to |o_1\rangle$ and $|e_2\rangle \to |o_2\rangle$ are allowed in the tunneling between $\gamma_1$ and the left lead, so we have

$$\sum_{\alpha,\beta} P_\alpha \Gamma^L_{\alpha\to\beta} = P_{e_1}\Gamma^L_{e_1\to o_1} + P_{e_2}\Gamma^L_{e_2\to o_2},$$
$$\sum_{\alpha,\beta} P_\beta \Gamma^L_{\beta\to\alpha} = P_{o_1}\Gamma^L_{o_1\to e_1} + P_{o_2}\Gamma^L_{o_2\to e_2}. \tag{S17}$$

At zero bias, using the formula of the differential conductance $G = \frac{\partial I}{\partial V}\big|_{V\to 0}$, the tunneling conductance reads as

$$G = \frac{e^2}{2T\hbar}\frac{\Gamma_L\Gamma_R}{\Gamma_L+\Gamma_R}\text{sech}(\frac{\nu}{T})^2\text{sech}[\frac{E_c(1-2\delta_g)}{2T}]^2. \tag{S18}$$

For Majorana case with level spacing $2\nu$, the analytical method is the same, while the impurity states in the island reduce to two eigenstates (even stste $|o\rangle$ and odd state $|o\rangle$). In the large temperautre limit $\nu/T \to 0$, we can easily obtain the expression of conductance for the standard Fu-teleportation [S5]

$$G_{Maj} = \frac{e^2}{2T\hbar}\frac{\Gamma_L\Gamma_R}{\Gamma_L+\Gamma_R}\text{sech}[\frac{2\nu + E_c(1-2\delta_g)}{2T}]^2. \tag{S19}$$

## V. 1E CONDUCTANCE OF THE DIRTY CASE IN $\nu \ll T \ll E_c$

Actually, the above calculation of the current is self-consistent and corresponds to a "clean" transport process, where the energy levels in the island are discrete and the island is only coupled with two leads. However, if the nanowire is also affected by disorder or the external environment, the self-consistency will be broken. In this case, the occupation of the energy levels is sophisticated and the entire island is more like in thermal equilibrium. We call this situation "dirty" transport. In this case, the occupation probabilities of different electronic states are proportional to the Fermi distribution function

$$P_{e_1} = P_0\frac{1}{e^{-\nu/kT}+1}, \quad P_{e_2} = P_0\frac{1}{e^{\nu/kT}+1},$$
$$P_{o_1} = P_1\frac{1}{e^{\nu/kT}+1}, \quad P_{o_2} = P_1\frac{1}{e^{-\nu/kT}+1}, \tag{S20}$$

where $P_{0/1}$ represents the probability to find the nanowire in the even/odd parity electronic state. In the steady state, the current satisfy

$$I = I_L = -I_R, \tag{S21}$$

where $I_L$ is expressed in Eq. (10), similarly $I_R$ is obtained by changing the superscript of the transition probability in $I_L$ from L to R

$$I_R = e \sum_{\alpha,\beta} P_\alpha \Gamma^R_{\alpha \to \beta} - e \sum_{\alpha,\beta} P_\beta \Gamma^R_{\beta \to \alpha}. \tag{S22}$$

From Eq. (S4), we have

$$\sum_{\alpha,\beta} P_\alpha \Gamma^R_{\alpha \to \beta} = P_{e_1} \Gamma^R_{e_1 \to o_2} + P_{e_2} \Gamma^R_{e_2 \to o_1},$$
$$\sum_{\alpha,\beta} P_\beta \Gamma^R_{\beta \to \alpha} = P_{o_1} \Gamma^R_{o_1 \to e_2} + P_{o_2} \Gamma^L_{o_2 \to e_1}. \tag{S23}$$

It is worth noting that although the expression of $I_R$ seems to be obtained only by changing the superscript of the transition probability in $I_L$ from L to R, however, because of the asymmetry, the transition of the impurity states has become different (i.e., left is $|e_1\rangle \to |o_1\rangle$ while right is $|e_1\rangle \to |o_2\rangle$).

Similarly, substituting Eq. (10) and Eq. (S22) into Eq. (S21), then combining normalization equation $P_0 + P_1 = 1$, one can figure out the differential conductance at zero-bias voltage

$$G = \frac{e^2 \Gamma_L \Gamma_R}{T\hbar}$$
$$\times \frac{\Gamma_L + 2\Gamma_R - 2\Gamma_R \cosh\left(\frac{\nu}{T}\right) + \Gamma_R \cosh\left(\frac{2\nu}{T}\right) + (\Gamma_L + \Gamma_R)\left[-1 + 2\cosh\left(\frac{\nu}{T}\right)\right] \cosh\left[\frac{E_c(1-2\delta_g)}{T}\right]}{\left(-\Gamma_L + 2\Gamma_L \cosh\left(\frac{\nu}{T}\right) + \Gamma_R \cosh\left(\frac{2\nu}{T}\right) + (\Gamma_L + \Gamma_R) \cosh\left[\frac{E_c(1-2\delta_g)}{T}\right]\right)^2}. \tag{S24}$$

At $\delta_g = 1/2$, the peak conductance becomes

$$G_{\text{peak}} = \frac{e^2}{2T\hbar} \frac{\Gamma_L \Gamma_R \operatorname{sech}\left(\frac{\nu}{T}\right)}{\Gamma_L + \Gamma_R \cosh\left(\frac{\nu}{T}\right)}, \tag{S25}$$



\end{document}